# The 'COVID' Crash of the 2020 U.S. Stock Market

Min Shu[1, *], Ruiqiang Song[2], Wei Zhu[3]

[1] Mathematics, Statistics & Computer Science Department, University of Wisconsin-Stout, Menomonie, WI, USA
[2] Michigan Technological University, Houghton, MI, USA
[3] Department of Applied Mathematics & Statistics, Stony Brook University, Stony Brook, NY, USA

## Abstract

We employed the log-periodic power law singularity (LPPLS) methodology to systematically investigate the 2020 stock market crash in the U.S. equities sectors with different levels of total market capitalizations through four major U.S. stock market indexes, including the Wilshire 5000 Total Market index, the S&P 500 index, the S&P MidCap 400 index, and the Russell 2000 index, representing the stocks overall, the large capitalization stocks, the middle capitalization stocks and the small capitalization stocks, respectively. During the 2020 U.S. stock market crash, all four indexes lost more than a third of their values within five weeks, while both the middle capitalization stocks and the small capitalization stocks have suffered much greater losses than the large capitalization stocks and stocks overall. Our results indicate that the price trajectories of these four stock market indexes prior to the 2020 stock market crash have clearly featured the obvious LPPLS bubble pattern and were indeed in a positive bubble regime. Contrary to the popular belief that the COVID-19 led to the 2020 stock market crash, the 2020 U.S. stock market crash was endogenous, stemming from the increasingly systemic instability of the stock market itself. We also performed the complementary post-mortem analysis of the 2020 U.S. stock market crash. Our analyses indicate that the probability density distributions of the critical time for these four indexes are positively skewed; the 2020 U.S. stock market crash originated from a bubble which began to form as early as September 2018; and the bubbles in stocks with different levels of total market capitalizations have significantly different starting time profiles. This study not only sheds new light on the making of the 2020 U.S. stock market crash but also creates a novel pipeline for future real-time crash detection and mechanism dissection of any financial market and/or economic index.

---

*Correspondence to: Mathematics, Statistics & Computer Science Department, 327 Jarvis Hall-Science Wing, University of Wisconsin-Stout, Menomonie, WI, USA.
*E-mail address:* shum@uwstout.edu (M. Shu), rsong1@mtu.edu (R. Song), wei.zhu@stonybrook.edu (W. Zhu)

# 1. Introduction

Until February 2020, the stock markets in the United States have experienced a 11-year boom period since the 2008 financial crisis, and the Wilshire 5000 Total Market index, widely accepted as the benchmark for the entire U.S. stock market, has increased dramatically by 403.5% from 6,858.4 on March 9, 2009 to 34,533.9 on February 19, 2020. However, starting on February 20, 2020, the U.S. stock market regime has suffered a sharp reversal from an upward trend to a downward trend. The Wilshire 5000 Total Market index dropped around 12,068.8 points in the next five weeks, which was a fall of 34.9 percent – its worst percentage loss in approximately one month since the Great Recession in 2008. During this crash, the S&P 500 index, one of the most followed stock indexes, dropped from 3,386.1 on February 19, 2020 to 2,237.4 on March 23, 2020, losing 33.9% of its value. The sharp drop of the S&P 500 index repeatedly triggered the level-1 trading curbs and caused major U.S. stock markets to suspend trading for 15 minutes on 3/9/2020, 3/12/2020, 3/16/2020, and 3/18/2020, respectively. Given that in the past 33 years, the trading curb has been triggered only once before on October 27, 1997, the consecutive four times triggering within 10 days are both historical and nerve wracking.

The 2020 U.S. Stock market Crash have greatly affected the lives and livelihoods of many people throughout the country, and caused permanent damage to the wealth of many investors, especially investors with little experience in risk management. During the crisis, the U.S. unemployment rate has quadrupled in just two months from 3.5% in February 2020 to 14.7% in April 2020 (FRED, 2020c).The total nonfarm payroll, which is a measure of the number of U.S. workers in the economy excluding proprietors, private household employees, unpaid volunteers, farm employees, and the unincorporated self-employed, and accounts for approximately 80 percent of the workers who contribute to GDP, dropped sharply from 152.463 million on February 2020 to 130.303 million on April 2020 (FRED, 2020b). Further, the yield of the US 10-year Treasury bond, one of the most popular debt instrument in the world and widely recognized as the lowest-risk investment because it is backed by the full faith and credit of the United States, fell sharply by 65.4% within three weeks from 1.56% on February 19, 2020 to 0.54% on March 9, 2020, its lowest close in history (FRED, 2020a). In order to stabilize the financial system and prevent the intensification of economic recession, the Federal Reserve, being the "lender of last resort", took the decisive action to rapidly reduce the target range of the federal funds rate to near zero. It has not only revived the Quantitative Easing (QE) program, but also expanded the QE purchases to an unlimited amount to support the flow of credit and hence bolster the economy (FED, 2020).

So far, the prevalent view is that the 2020 stock market crash was mainly caused by external shocks. The fiction is that the rapid spread of the novel coronavirus COVID-19 since January 2020 and the subsequent lockdowns have caused the crash (Albuquerque et al., 2020), and therefore; the 2020 Stock Market Crash is also called the Great Coronavirus Crash (Coy, 2020). In addition to the COVID-19 pandemic, the mammoth nonfinancial corporate debt bulge (Lynch, 2020) and the 2020 Russia–Saudi Arabia oil price war (Guardian, 2020) are also thought to have triggered the 2020 stock market crash.

In this research, we employed the Log Periodic Power Law Singularity (LPPLS) model (Drozdz et al., 1999; Feigenbaum & Freund, 1996; Johansen et al., 2000; Johansen et al., 1999; Sornette

& Johansen, 2001; Sornette et al., 1996) to systematically analyze the 2020 stock market crash in the United States. A fusion of statistical physics, financial economics, and behavioral finance, the LPPLS model syndicates the mathematical and statistical physics of bifurcations and phase transitions, the economic theory of rational expectations, and behavioral finance of herding of traders to model a positive (or negative) financial bubbles as a process of unsustainably faster-than-exponential growth (or drop) to reach an infinite return in finite time, leading to a short-term correction molded by the symmetry of discrete scale invariance (Sornette, 1998). In the LPPLS model, there are two types of agents: the rational traders who trade based on rational expectations, and the noise traders who are prone to exhibit imitation and herding behavior. Given that asset prices can be destabilized by the collective behavior of noise traders through the massive herding and imitation transactions, the LPPLS model diagnose financial bubbles by capturing two distinct characteristics of price trajectories normally observed in bubble regimes: the faster-than-exponential growth caused by positive feedbacks by imitation and herding behavior of noise traders, and the accelerating log-periodic volatility fluctuations of the price growth due to expectations of higher returns and an upcoming crash.

The LPPLS model has become an increasingly recognized tool to diagnose speculative bubbles based on its characteristic patterns embedded in asset price trajectories. A quick review of its recent history ensues. Filimonov and Sornette (2013) simplified its calibration by transforming the traditional LPPLS formula reducing the nonlinear parameters from four to three. Sornette et al. (2015) proposed the LPPLS Confidence indicator and Trust indicator to qualify the sensitivity of the observed bubble pattern. Filimonov et al. (2017) calibrated the LPPLS model via the adjusted profile likelihood. Demos and Sornette (2017) quantified the uncertainty of the calibrated bubble start and end times using the eigenvalues of the Hessian matrix. Demirer et al. (2019) presented the LPPLS confidence multi-scale indicators to study the predictive power of market-based indicators. Shu and Zhu (2020b) proposed an adaptive multilevel time series detection methodology to detect the real-time status of financial bubbles and crashes. In addition, the LPPLS model has been broadly deployed to diagnose speculative bubbles and crashes in various financial markets including the stock market (Demirer et al., 2019; Demos & Sornette, 2017; Filimonov et al., 2017; Jiang et al., 2010; Li, 2017; Shu, 2019; Shu & Zhu, 2019, 2020a; Song et al., 2021; Sornette et al., 2015; Yan et al., 2010; Zhang et al., 2016), the cryptocurrency market (Gerlach et al., 2019; Shu & Zhu, 2020b), the real estate market (Zhou & Sornette, 2003, 2006, 2008), and the energy market (Sornette et al., 2009; Zhang & Yao, 2016). It should be pointed out that the LPPLS model can only diagnose endogenous bubbles where the changes in price trajectories are caused by the self-reinforcing synergistic herding and imitative behaviors through interactions of market participants in the long-memory process of endogenous organizations. A crash observed but not deteced by the LPPLS model is likely an exogenous (or external) crash caused by changes of fundamental values of assets resulting from exogenous shocks. For exogenous crashes, the price trajectories would not present the two distinct characteristics of endogenous bubbles, that is, the faster-than-exponential growth and the accelerating log-periodic volatility fluctuations of the price growth when approaching to finite time singularity.

In order to investigate the 2020 stock market crash in U.S. equities sectors with different levels of total market capitalization, we used four major U.S. stock market indexes, including the Wilshire 5000 Total Market (W5000) index, the S&P 500 (SP500) index, the S&P MidCap 400 (SP400) index, and the Russell 2000 (R2000) index. The W5000 index measures the overall performance of most publicly traded U.S. headquartered equities and is widely accepted as the definitive benchmark for the market value of all stocks actively traded in the United States. The SP500 index, maintained by the S&P Dow Jones Indices, measures the stock performance of 500 large capitalization stocks listed on the U.S. stock exchanges and is the most commonly tracked stock index. The SP400 index, also maintained by the S&P Dow Jones Indices, is the most followed benchmarks for middle capitalization stocks. The R2000 index, maintained by the FTSE Russell, measures the performance of the smallest 2,000 stocks in the Russell 3000 Index and serves as a barometer for the small capitalization stocks. These four indexes are all market-capitalization-weighted stock market indexes. Figure 1 shows the evolution of their price trajectories from January 2015 to December 2020.

In this study, we detected the positive and negative bubbles in the U.S. stock market based on the LPPLS model using the daily data of the W5000, SP500, SP400, and R2000 indexes from January 2019 through December 2020 to investigate the 2020 U.S. stock market crash. This study also presented a complementary post-mortem analysis on the nature of the crash. This paper is laid out as the following: Section 2 presents the methodology used including the LPPLS model calibration and confidence indicator. Section 3 discusses the empirical analysis and results; finally, Section 4 recaps.

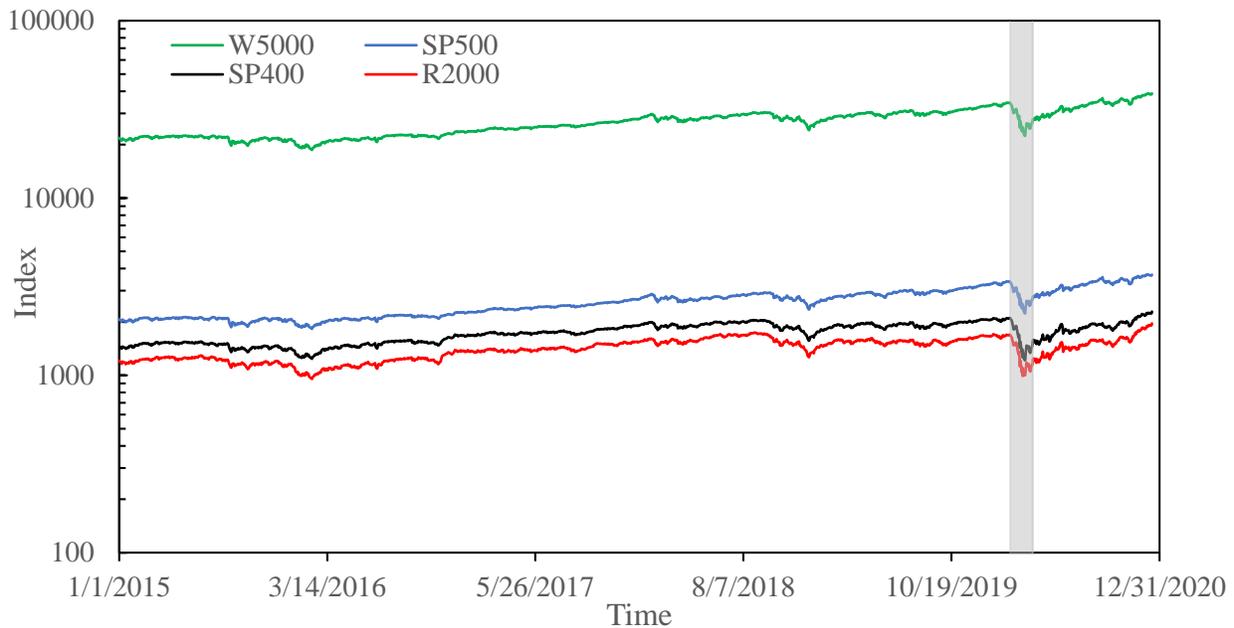

Figure 1. Evolution of price trajectories of the W5000, SP500, SP400, and R2000 indexes from January 2015 to December 2020. The shadowed band shows the period of the 2020 U.S. stock market crash.

## 2. Methodology

### 2.1 The Log-Periodic Power Law Singularity (LPPLS) Model

The LPPLS model is also called the Johansen–Ledoit–Sornette (JLS) model or Log-Periodic Power Law (LPPL) Model. Based on the assumption that the observed price trajectory of an asset during a bubble period deviates from its intrinsic fundamental value (Kindleberger & Aliber, 2011; Sornette, 2017), the LPPLS model combines two typical bubble features: the transient super-exponential growth and the accelerating log-periodic volatility fluctuations, to model the price trajectory in a bubble regime as a process of fast-than-exponential power law growth punctuated by short-lived corrections. In its core the LPPLS model is a parametric method highly efficient in bubble detections and can outperform machine learning algorithms that are less parametric and more data driven – and hence less agile and sensitive when the data volume is modest, which is often the case for financial times series analysis. The dynamics of the observed asset price $p(t)$ in the LPPLS model is assumed to follow (Johansen et al., 2000):

$$\frac{dp}{p} = \mu(t)dt + \sigma(t)dW - kdj \qquad (1)$$

where $\mu(t)$ is the time-dependent expected return, $\sigma(t)$ is the volatility, $dW$ is the infinitesimal increment of a standard Wiener process, $dj$ represents a discontinuous jump with the value of $j=0$ before and $j=1$ after the crash happens, while $k$ measures the amplitude of a potential crash. The crash hazard rate $h(t)$ quantifying the crash probability at a specified time $t$ controls dynamics of the jumps. The expected value of $dj$ between $t$ and $t + dt$ on the condition that no jump has yet happened can be determined as $E[dj] = h(t)\mathrm{d}t$. The LPPLS model reflects the joint action of two types of agents: (a) rational traders with rational expectations and, (b) noise traders who are susceptible to exhibit herding behavior and imitation, thereby making irrational decisions of buying, selling or holding. On the assumption that the asset prices can be destabilized by the collective behavior of noise traders, the aggregate effect of noise traders can be determined by the dynamics of the crash hazard rate $h(t)$ during the course a bubble (Johansen et al., 2000):

$$h(t) = \alpha(t_c - t)^{m-1}(1 + \beta cos(\omega \ln(t_c - t) - \phi')) \qquad (2)$$

where $\alpha, \beta, \omega,$ and $t_c$ are the model parameters. For a given time $t$, the crash risk is a sum of the power law singularity $\alpha(t_c - t)^{m-1}$ which embodies the positive feedback mechanism resulting from the herding behaviors of the noise traders, decked by large scale amplitude periodic oscillations in the logarithm of the time to the critical time $t_c$. The log periodic function $\cos(\omega \ln(t_c - t) - \phi')$ reflects a potential hierarchical cascade of panic acceleration leading to punctuate the growth of price trajectories. The log-periodicity can be caused by the preexisting hierarchy in noise trader sizes (Sornette & Johansen, 1997) and/or from the interplay between market price impact inertia and nonlinear intrinsic fundamental value investing (Ide & Sornette, 2002).

Under the non-arbitrage and rational expectation conditions, the unconditional expectation $E[dp]$ of the price increment should be zero, leading to:

$$\mu(t) \equiv E[\frac{dp/dt}{p}]_{\text{no crash}} = kh(t) \tag{3}$$

thus, the expected return $\mu(t)$ is proportional to the crash hazard rate $h(t)$, so that the higher the crash risk, the higher the expected return, and vice versa. If no crash has yet occurred in a bubble regime, the simple mathematical formula of the LPPLS for the expected trajectory of the log-price can be derived by solving Equations (1), (2) and (3) simultaneously (Sornette, 2003):

$$\text{LPPLS}(t) \equiv \ln E[p(t)] = A + B(t_c - t)^m + C(t_c - t)^m \cos[\omega \ln(t_c - t) - \phi] \tag{4}$$

where $B = -k\alpha/m$ and $C = -k\alpha\beta/\sqrt{m^2 + \omega^2}$. The term $C\cos[.]$ can be extended with two linear parameters $C_1 = C\cos\phi$ and $C_2 = C\sin\phi$ to replace the linear parameter $C$ and the nonlinear parameter $\phi$. Then the simple mathematical formula of the LPPLS in Equation (4) can be rephrased as (Filimonov & Sornette, 2013):

$$\begin{aligned}\text{LPPLS}(t) \equiv E_t[\ln p(t)] = A &+ B(t_c - t)^m + C_1(t_c - t)^m \cos[\omega \ln(t_c - t)] \\ &+ C_2(t_c - t)^m \sin[\omega \ln(t_c - t)]\end{aligned} \tag{5}$$

where $A > 0$ is the expected value of the log-price at the critical time $t_c$. In a bubble regime, the power parameter $m$ ranges between 0 and 1 to ensure that not only the price remains finite at the $t_c$, but also the expected logarithmic price diverges at the $t_c$. The critical time $t_c$ is the most probable time for the asset price trajectory to undergo a regime change in a form of a major crash or a great change of growth rate with terminating the accelerated oscillations.

## 2.2 Model Calibration and LPPLS Confidence Indicator

The Ordinary Least Squares method is used to calibrate the LPPLS model and estimate the thee nonlinear parameters $\{t_c, m, \omega\}$ and the four linear parameters $\{A, B, C_1, C_2\}$. Based on the $L^2$ norm, the sum of squares of residuals of the converted LPPLS formula in Equation (5) can be written as:

$$F(t_c, m, \omega, A, B, C_1, C_2) = \sum_{i=1}^{N} [\ln p(\tau_i) - A - B(t_c - \tau_i)^m - C_1(t_c - \tau_i)^m \cos(\omega \ln(t_c - \tau_i)) \\ - C_2(t_c - \tau_i)^m \sin(\omega \ln(t_c - \tau_i))]^2 \tag{6}$$

where $\tau_1 = t_1$ and $\tau_N = t_2$. Subordinating the four linear parameters $(A, B, C_1, C_2)$ to the three nonlinear parameters $(t_c, m, \omega)$, we can determine the cost function $\chi^2(t_c, m, \omega)$:

$$\chi^2(t_c, m, \omega) = F_1(t_c, m, \omega) = \min_{\{A,B,C_1,C_2\}} F(t_c, m, \omega, A, B, C_1, C_2) = F(t_c, m, \omega, \hat{A}, \hat{B}, \hat{C}_1, \hat{C}_2) \tag{7}$$

where the hat symbol ^ indicates the estimated parameters. The three nonlinear parameters $\{t_c, m, \omega\}$ can be estimated by solving the nonlinear optimization equation:

$$\{\hat{t}_c, \hat{m}, \hat{\omega}\} = arg \min_{\{t_c, m, \omega\}} F_1(t_c, m, \omega) \qquad (8)$$

The four linear parameters $(A, B, C_1, C_2)$ can be obtained by solving the following optimization equation:

$$(\hat{A}, \hat{B}, \hat{C}_1, \hat{C}_2) = arg \min_{(A,B,C_1,C_2)} F(t_c, m, \omega, A, B, C_1, C_2) \qquad (9)$$

Equation (9) can be solved analytically through the matrix equations:

$$\begin{pmatrix} N & \sum f_i & \sum g_i & \sum h_i \\ \sum f_i & \sum f^2_i & \sum f_i g_i & \sum f_i h_i \\ \sum g_i & \sum f_i g_i & \sum g^2_i & \sum h_i g_i \\ \sum h_i & \sum f_i h_i & \sum g_i h_i & \sum h^2_i \end{pmatrix} \begin{pmatrix} \hat{A} \\ \hat{B} \\ \hat{C}_1 \\ \hat{C}_2 \end{pmatrix} = \begin{pmatrix} \sum \ln p_i \\ \sum f_i \ln p_i \\ \sum g_i \ln p_i \\ \sum h_i \ln p_i \end{pmatrix} \qquad (10)$$

where $f_i = (t_c - t_i)^m$, $g_i = (t_c - t_i)^m cos(\omega \ln(t_c - t_i))$, and $h_i = (t_c - t_i)^m sin(\omega \ln(t_c - t_i))$.

Here the covariance matrix adaptation evolution strategy (CMA-ES) developed by Hansen et al. (1995) was employed towards the optimization. The three nonlinear parameters $(t_c, m, \omega)$ can be estimated by minimizing the sum of residuals between the fitted LPPLS model and the observed price trajectory. To minimize calibration problems of the LPPL model, the search space is set to (Shu & Zhu, 2020b):

$$m \in [0,1], \omega \in [1, 50], t_c \in \left[t_2, t_2 + \frac{t_2 - t_1}{3}\right], \frac{m|B|}{\omega \sqrt{C_1^2 + C_2^2}} \geq 1 \qquad (11)$$

The condition $t_c \in [t_2, t_2 + (t_2 - t_1)/3]$ ensures that the fitted critical time $t_c$ falls beyond the endpoint $t_2$, but adjacent to the $t_2$ since the LPPLS model has a degraded predictive capacity far beyond $t_2$ (Jiang et al., 2010). Here, the Damping parameter $m|B|/\left(\omega \sqrt{C_1^2 + C_2^2}\right) \geq 1$ because the crash hazard rate $h(t)$ is defined as a non-negative value (Bothmer & Meister, 2003).

Furthermore, to select the valid calibration results and ensure model rigorousness (Brée et al., 2013; Sornette et al., 2013), the fitted LPPLS model results are filtered under the conditions:

$$m \in [0.01, 0.99], \omega \in [2, 25], t_c \in \left[t_2, t_2 + \frac{t_2 - t_1}{5}\right], \frac{\omega}{2} \ln\left(\frac{t_c - t_1}{t_c - t_2}\right) \geq 2.5,$$
$$\max\left(\frac{|\hat{p}_t - p_t|}{p_t}\right) \leq 0.20, p_{lomb} \leq \alpha_{sign}, \ln(\hat{p}_t) - \ln(p_t) \sim AR(1) \qquad (12)$$

These filters conditions are derived from empirical evidence gathered in previous investigations on financial bubbles (Jiang et al., 2010; Sornette et al., 2015). The $(\omega/\pi)\ln[(t_c - t_1)/(t_c - t_2)] \geq 2.5$ condition is employed to distinguish between genuine log-periodic signals and noise-generated signals (Huang et al., 2000). The $\max(|\hat{p}_t - p_t|/p_t) \leq 0.20$ condition ensures that the calibrated price $\hat{p}_t$ is close to the actual asset price $p_t$ (Shu & Zhu, 2020b). The $P_{lomb} \leq \alpha_{sig}$ condition applies the Lomb spectral analysis to the time series of the detrended residuals $r(t) = (t_c - t)^{-m}(\ln[p(t)] - A - B(t_c - t)^m)$ to ensure the existence the logarithm-periodic oscillations in the fitted model (Sornette & Zhou, 2002). The $\ln(\hat{p}_t) - \ln(p_t) \sim AR(1)$ condition ensures that the logarithmic price in a bubble stage is due to a deterministic LPPLS component, the LPPLS fitting residuals can be constructed through the mean-reversal Ornstein-Uhlenbeck (O-U) process (Lin et al., 2014). Only calibration results meeting the filter constrains in Equation (12) are considered valid, while the remaining results are deemed superfluous.

To measure the sensitivity of the calibration results with respect to the selection of the start time $t_1$ for the fitting windows, the LPPLS confidence indicator proposed by Sornette et al. (2015) is utilized. It is defined as the proportion of the fitting windows for the calibrated LPPLS model that meets the specified filter conditions. Larger LPPLS confidence indicator means more fitting windows have the detected bubble pattern of the LPPLS model, and hence more reliable implication of price trajectories in a bubble phrase. In contrast, the smaller LPPLS confidence indicator means the less fitting windows would present the LPPLS bubble pattern, and therefore more fragility of the bubble signals.

For a given fictitious "present" data point $t_2$, the LPPLS confidence indicator is calculated in the following steps: (i) create a series of fitting time windows by moving the start time $t_1$ toward the $t_2$ with a step of $dt$, (ii) calibrate the model within the given search space for each fitting time window, count the number of fitting time windows that meets the specified filter constrains, and (iii) divide this number by the total number of the fitting windows. The LPPLS confidence indicator is causal because only data prior to $t_2$ are used in the calculation of the LPPLS confidence indicator.

## 3. Empirical analysis

### 3.1 LPPLS bubble identification

In this section, we adopted the LPPLS confidence indicator as a diagnostic tool to detect both positive and negative bubbles in the U.S. stock markets based on the daily data of the W5000, SP500, SP400, and R2000 indexes from January 2, 2019 to December 15, 2020. Data on stock market indexes came from Yahoo Finance (https://finance.yahoo.com/). A positive bubble is associated with the accelerating growth trend which is susceptible to changes of regime in the form of large crashes or volatile sideway plateaus, while a negative bubble is related to the accelerating downward trend which is vulnerable to regime changes in the form of rallies or volatile sideway plateaus. In this study, the LPPLS confidence indicator is calculated by shrinking the length of time windows $t_2 - t_1$ from 650 trading days to 30 trading days in steps of 5 trading days, and moving the endpoint $t_2$ from January 2, 2019 to December 15, 2020.

Figure 2 shows the results of the LPPLS confidence indicator for these four US stock market indexes, in which the positive bubbles is shown in red and negative bubbles in green (right scale) along with the index price in blue (left scale). From Figure 2, we can perceive intuitively the confidence level of the detected LPPLS bubble pattern in the price trajectories because the confidence indicator measures the sensitivity of the fitting results with respect to the start time selection. When the a substantial number of the calibrated results with different start time points can pass the filter conditions in Equation (12), the value of the confidence indicator for a specified pseudo-present time can reach a high enough value such as 10%, indicating that the detected LPPLS signature is relatively robust to the selection of the start time, and therefore the price trajectory can be confirmed in bubble regime. On the contrary, when most of the calibrated results with different start time points fail to pass the filter conditions, the value of the confidence indicator can be low close to zero, indicating that the detected LPPLS signature is highly sensitive to the choice of the start times and there the price trajectory is unlikely to be in a true bubble regime.

The detected bubble status of W5000 index is shown in Figure 2 (a), including four obvious clusters of positive bubbles between December 13, 2019 and January 28, 2020, between February 11 and March 4, 2020, between August 10 and September 4, 2020, and between November 23 and December 15, 2020. There is also one subtle cluster of negative bubble between March 26 and March 27, 2020. On February 20, 2020, the positive LPPLS confidence indicator reached the global peak value of 11.2%, that is, 14 out of 125 fitting windows can successfully pass the filter, indicating that the detected LPPLS bubble pattern is reliable and the price trajectory of W5000 index can be confirmed in a positive bubble regime. There is a strong possibility that the accelerated growth trend of the W5000 index is unsustainable, and the positive bubble regime of the W5000 index is likely to change. The forecast of bubble regime change was confirmed by the fact that the W5000 plunged dramatically from 34,533.9 on February 19, 2020 to 22,465.1 on March 23, 2020, losing 34.9% of its value within 24 trading days. During the 2020 U.S. stock market crash, similar detected positive bubble patterns shown in Figure 2 (a) for the W5000 index can also be found in the remaining subfigures in Figure 2, including: (b) SP500, (c) SP400, and (d) R2000. Furthermore, the negative LPPLS confidence indicators for W5000 peaked at the value of 1.6% on March 27, 2020, for SP500 peaked at 1.6% on March 26, 2020, for SP400 peaked at 0.8% on March 19, 2020, and for R2000 peaked at 1.6% on March 24, 2020, respectively. The subtle clusters of negative LPPLS confidence indicators for these four indexes indicate that there is no obvious negative LPPLS bubble pattern presented in the price trajectories during the 2020 U.S. stock market crash.

The statistics of positive bubble detection results for these four stock market indexes during the 2020 U.S. stock market crash and the related information of the peaks and valleys are summarized in Table 1. During the 2020 stock market crash, all four indexes lost more than a third of their values within five weeks. The SP400 index has the largest crash size of 42.1% dropping from 2,106.1 on February 20, 2020 to 1,218.6 on March 23, 2020, followed by the R2000 index with a crash of 41.6% from 1,696.1 on February 20, 2020 to 991.2 on March 23, 2020, while the SP500 index has the smallest crash size of 33.9% dropping from 3,386.1 on February 19, 2020 to 2,237.4 on March 23, 2020, indicating that both middle capitalization stocks and small capitalization stocks have suffered much greater losses than the large capitalization stocks. Table 1 lists the peak values of the LPPLS confidence indicator (CI) during the 2020 U.S. stock market crash. The SP500 index has the largest peak CI value of 16.0%, meaning that 20 out of 125 fitting windows can successfully meet the filter constrains and thus strong LPPLS bubble signal has presented in the price trajectory of the SP500 index in the 2020 U.S. stock market crash. These four indexes have the peak confidence indicator values exceeding 6.4%, indicating that the price trajectories of these four stock market indexes clearly present the LPPLS bubble pattern and are indeed in a positive bubble regime. Because the LPPLS model can only detect the endogenous bubbles, the postive bubbles as well as the subsequent crashes in these four indexes during the 2020 U.S. stock market crash are endogenous stemming from the increasingly systemic instability of the stock market itself, while the well-known external shocks, such as the COVID-19 pandemic-induced market instability, the mass hysteria, and the corporate debt bubble, are not the root causes of the 2020 U.S. stock market crash, and they only served as sparks during the 2020 stock market crash.

Table 1: Statistics of positive bubble detection based on daily data during the 2020 U.S. stock market crash

| Index | Peak Day | Peak Price | Valley Date | Valley Price | Crash Size | Peak CI |
|---|---|---|---|---|---|---|
| W5000 | 2/19/2020 | 34533.9 | 3/23/2020 | 22465.1 | 34.9% | 11.2% |
| SP500 | 2/19/2020 | 3386.1 | 3/23/2020 | 2237.4 | 33.9% | 16.0% |
| SP400 | 2/20/2020 | 2106.1 | 3/23/2020 | 1218.6 | 42.1% | 7.2% |
| R2000 | 2/20/2020 | 1696.1 | 3/18/2020 | 991.2 | 41.6% | 6.4% |

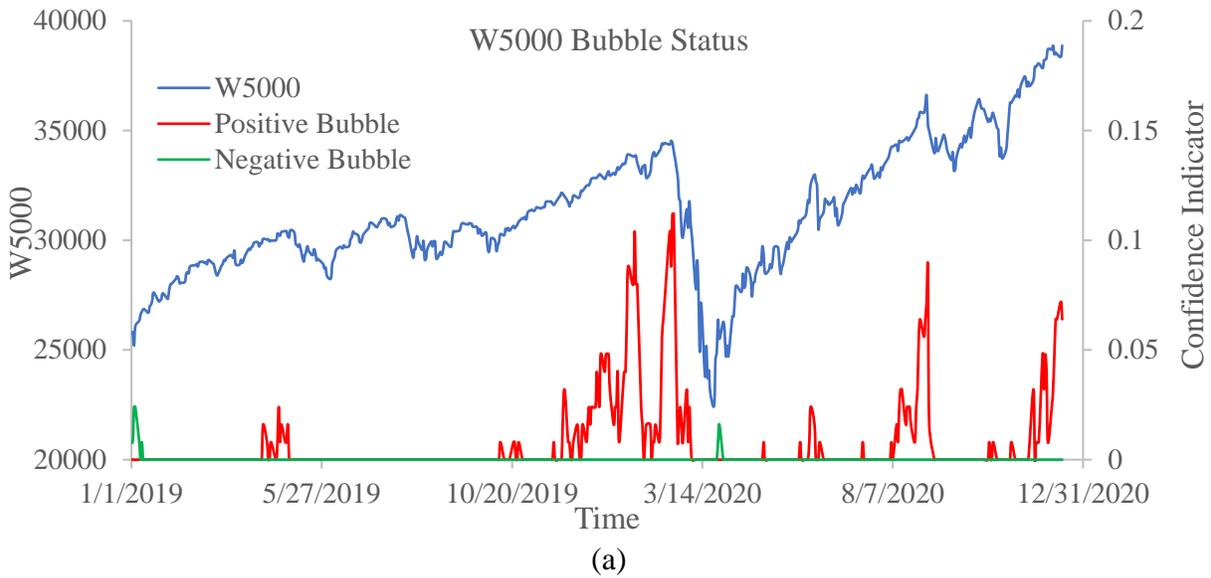

(a)

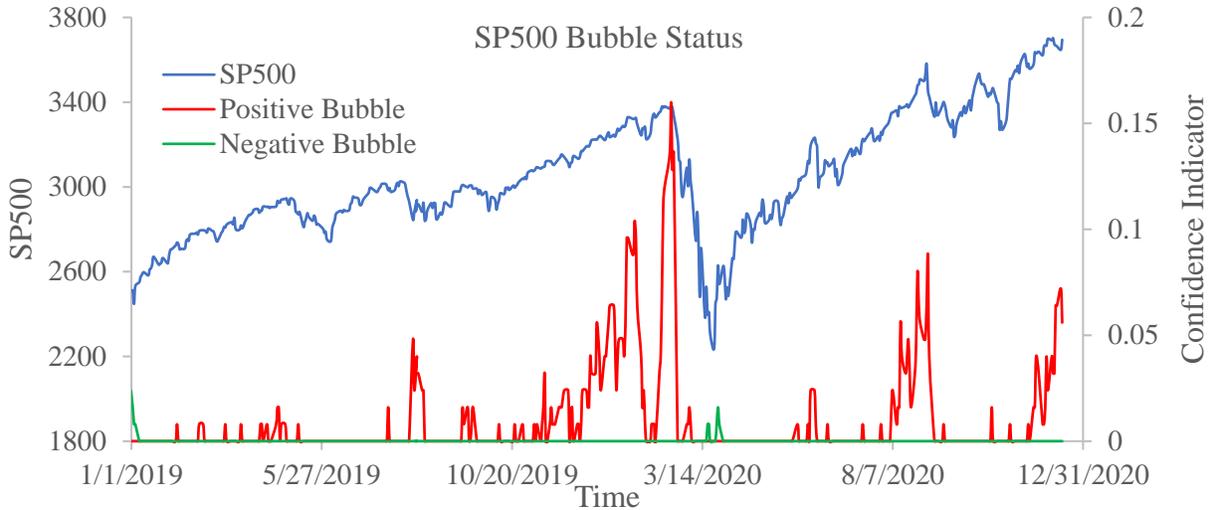

(b)

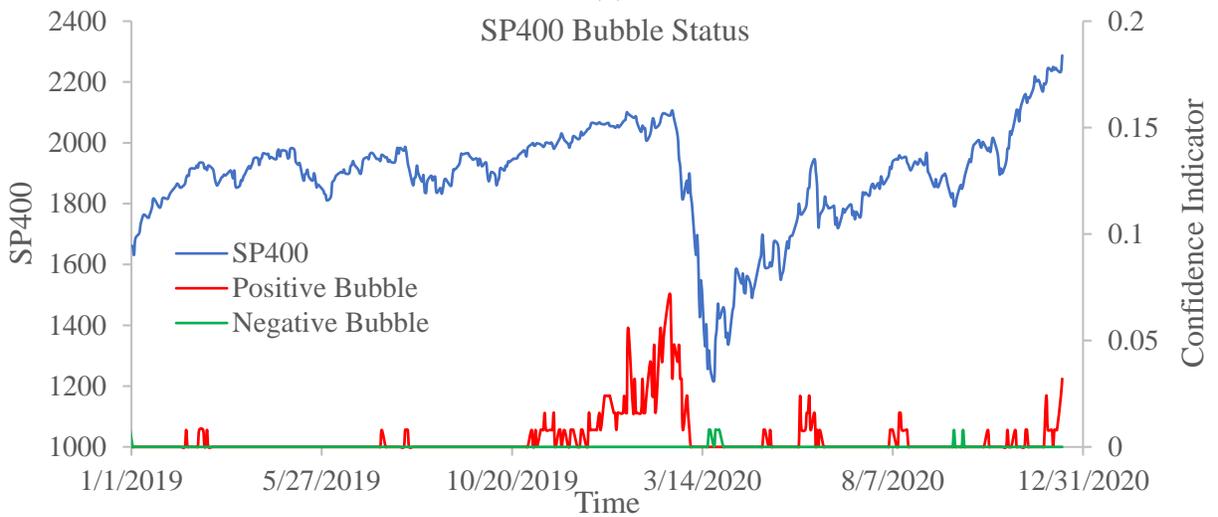

(c)

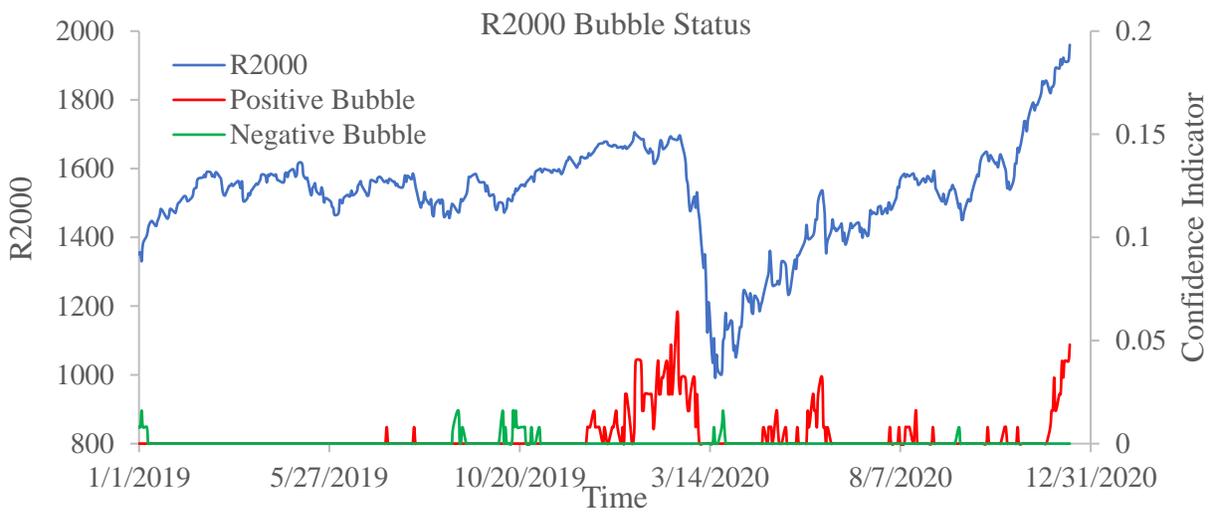

(d)

Figure 2. LPPLS confidence indicator for positive bubbles is shown in red and negative bubbles in green (right scale) along with the index price in blue (left scale) for the four U.S. stock market indexes based on daily data from January 2019 to December 2020.

## 3.2 Post-mortem analysis for the bubbles

This section presents the complementary post-mortem analysis to provide additional information on the estimated LPPLS models for diagnosing the 2020 U.S. stock market crash. The calibrated results that successfully pass the filter conditions in Equation (12) with endpoints falling in the positive bubble clusters during the 2020 U.S. stock market crash were collected to perform the post-mortem analysis on the four indexes: W5000, SP500, SP400, and R2000. In order to create the probability density distribution of the predicted critical time $t_c$ and the start time points $t_1$, the Kernel density estimation (Davis et al., 2011; Parzen, 1962) are adopted in this study.

Figure 3 shows the probability density distribution of the predicted critical time $t_c$'s as well as the estimated beginning time $t_1$'s for the four U.S. stock market indexes during the 2020 U.S. stock market crash. The optimal values for the bubble starting date $t_1$ are represented by the probability density distribution $pdf(t_1)$ in green, while the forecasted critical time $t_c$ is depicted by the probability density distribution $pdf(t_c)$ in red which can be viewed as the probability measure of the time of crash. Three typical fitting examples with different time scale windows are shown in Figure 3 for each index.

In particular, the results of the post-mortem analysis for the W5000 index is shown in Figure 3 (a). There are 99 calibrated results collected in the positive cluster during the 2020 U.S. stock market crash. The ranges of the start time point $t_1$ and the end time point $t_2$ for the 99 selected fitting windows are from September 20, 2018 to November 8, 2019 and from February 7, 2020 to March 4, 2020, respectively. It can be observed that the $pdf(t_1)$ is concentrated on the time frame when the W5000 index begins to super-exponentially increase and the beginning of the W5000 index bubble can be determined as early as September 20, 2018. As shown in Figure 3 (a), the 20%/80% and 5%/95% quantile range of the crash dates $t_c$ for the W5000 index during the 2020 U.S. stock market crash are from February 18, 2020 to March 27, 2020, and from February 12, 2020 to May 13, 2020, respectively. The observed peak date for the W5000 index is February 19, 2020, which lies within the quantile ranges of the predicted crash dates $t_c$ and the $pdf(t_c)$ is positively skewed with the peak value on February 20, 2020. Figure 3 (a) also presents three typical fitting examples with a common endpoint $t_2$= February 14, 2020, and three different sets of starting time: $t_1$= September 24, 2018, $t_1$= July 26, 2019, and $t_1$= October 28, 2019, --- corresponding to the long, middle and short time scale windows, respectively.

The post-mortem analysis results for the SP500, SP400 and R2000 indexes, are presented in Figure 3 (b), (c) and (d), respectively. The 20%/80% quantile range of the crash dates $t_c$ for the SP500, SP400 and R2000 indexes during the 2020 U.S. stock market crash are from February 18 to April 3, 2020, from January 21 to April 1, 2020, and from January 28 to February 26, 2020, respectively. It should be noted that all four probability density distributions of the predicted critical time $t_c$ are positively skewed and the mass of distributions are concentrated on the right side of the figures. In addition, all four indexes have the minimum start time points $t_1$ on September 2018, indicating that the 2020 U.S. stock market crash originated from a bubble which began to form as early as September 2018. Furthermore, the probability density distributions of the start time point $t_1$ can be seen not all equal for these four indexes. The W5000 index has a similar $pdf(t_1)$ shape as the SP500 index. However, the SP400 index and R2000 index have the distinctly different $pdf(t_1)$ shapes compared with the W5000 and SP500 indexes, signifying that the bubbles in stocks with different levels of total market capitalizations may have significantly different starting time profiles.

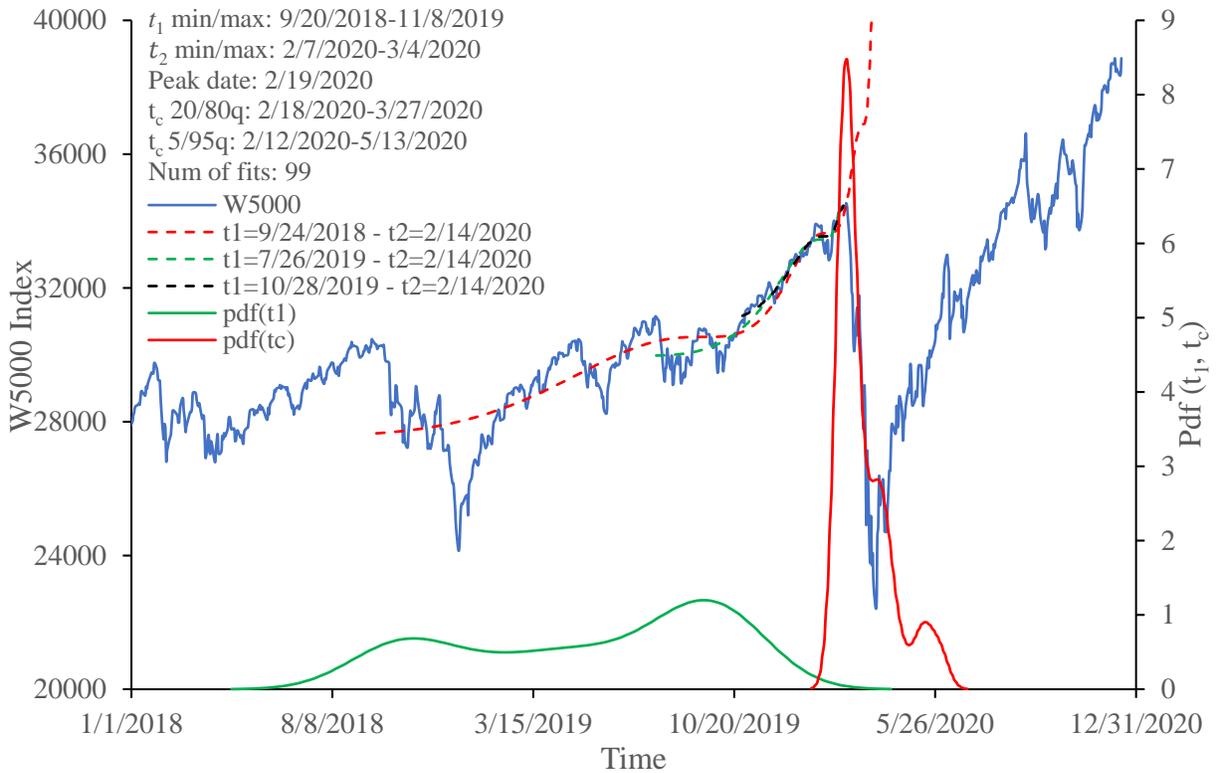

(a) Post-mortem analysis for W5000 index

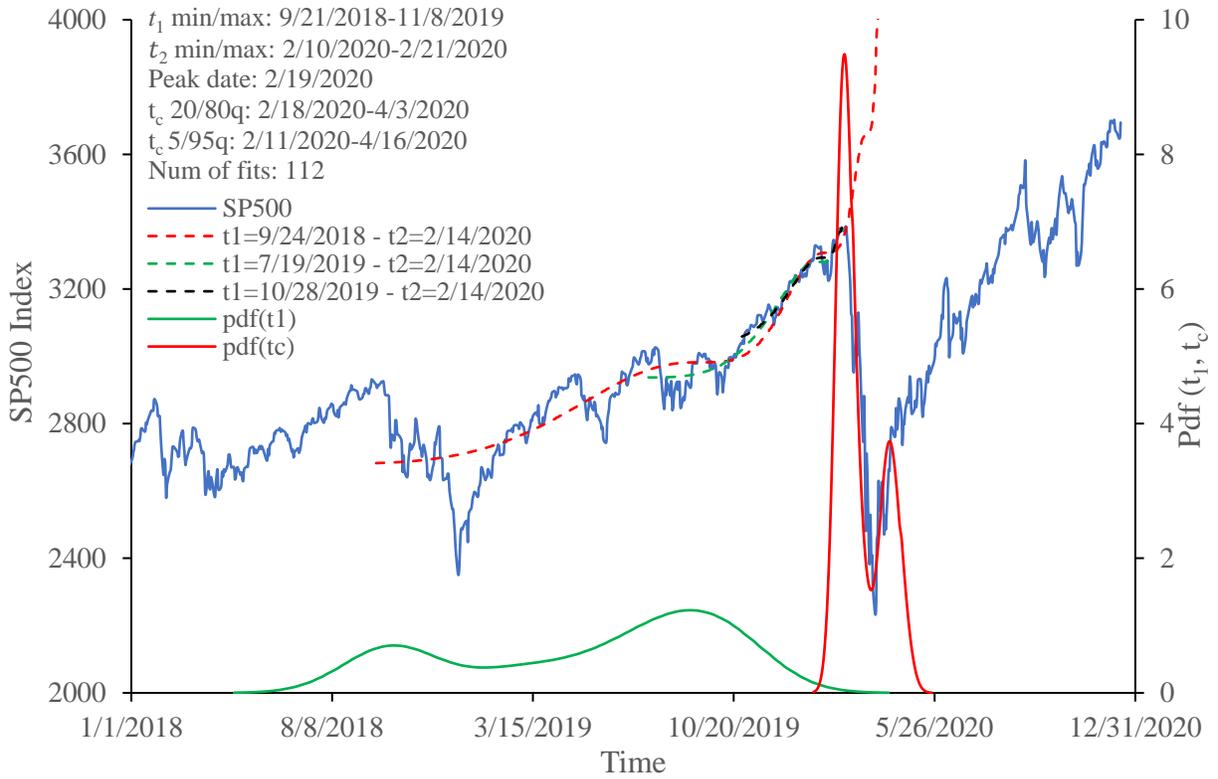

(b) Post-mortem analysis for SP500 index

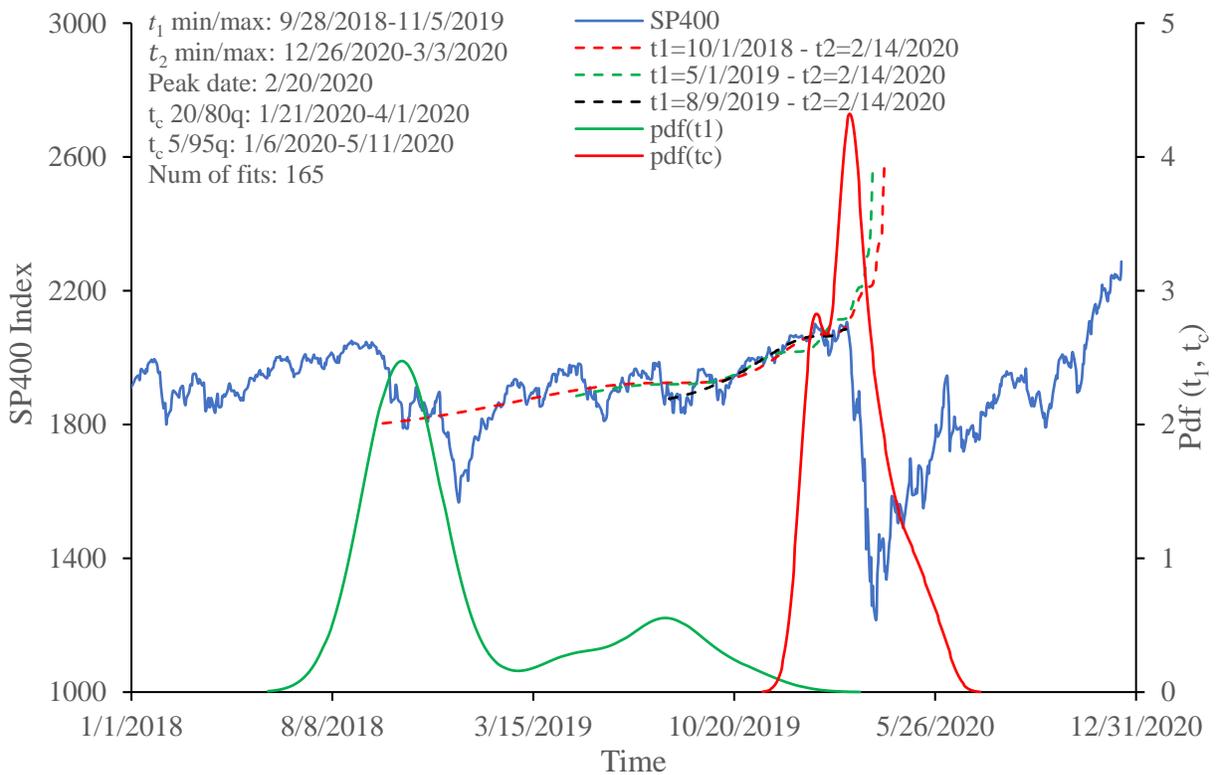

(c) Post-mortem analysis for SP400 index

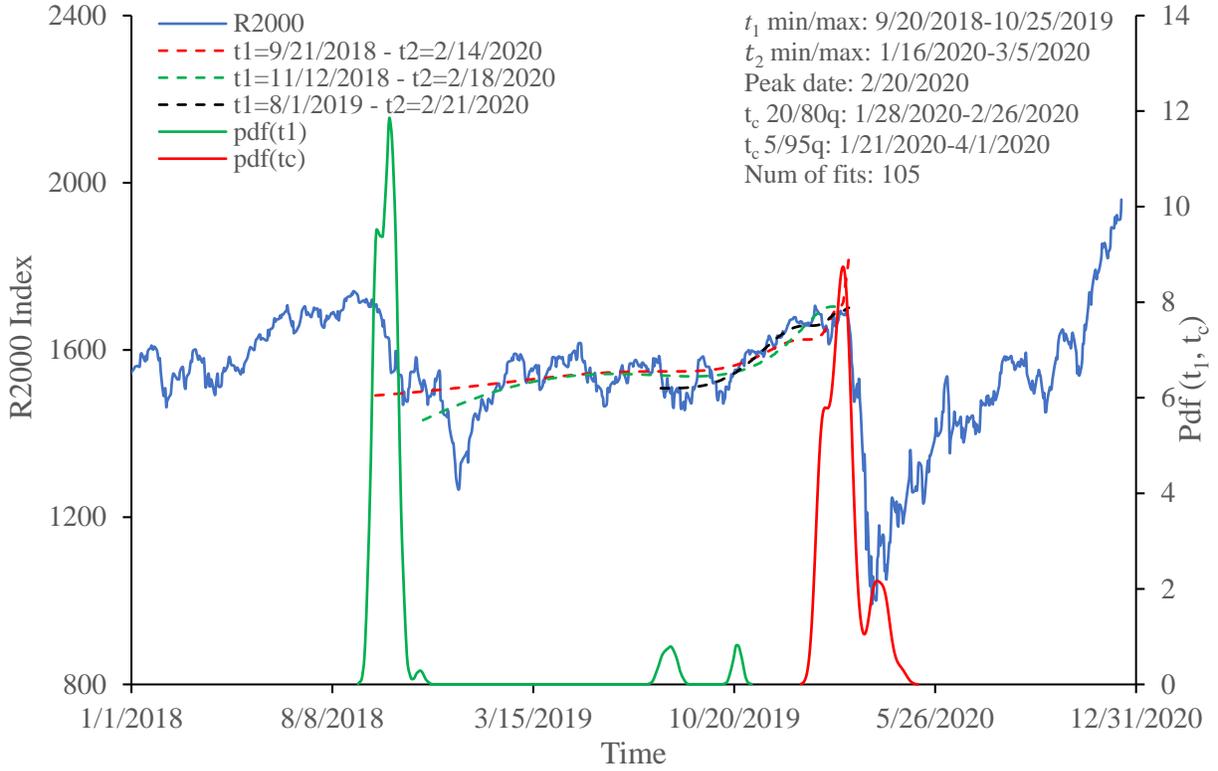

(d) Post-mortem analysis for R2000 index

Figure 3. The probability density distributions $pdf\,(t_1, t_c)$ for the 2020 U.S. Stock market Crash (right scale) together with the index price in blue (left scale) for the four U.S. stock market indexes based on daily data from January 2018 to December 2020.

## 4. Conclusions

We employed the LPPLS methodology in this study to systematically dissect the 2020 stock market crash in the U.S. equities sectors with different levels of total market capitalizations through four major U.S. stock market indexes, including the W5000, SP500, SP400 and R2000, representing the stocks overall, the large capitalization stocks, the middle capitalization stocks and the small capitalization stocks, respectively. During the 2020 U.S. stock market crash, all four indexes lost more than a third of their values within five weeks, while both middle capitalization stocks and small capitalization stocks have suffered much greater losses than the large capitalization stocks and stocks overall. Our results indicate that the price trajectories of these four stock market indexes prior to 2020 stock market crash have clearly featured the obvious LPPLS bubble pattern of the faster-than-exponential growth corrected by the accelerating logarithm-periodic oscillations and were indeed in a positive bubble regime. Contrary to the popular belief that the COVID-19 pandemic-induced market instability led to the 2020 U.S. stock market crash, the crashes in these four indexes during the 2020 U.S. stock market crash were endogenous stemming from the increasingly systemic instability of the stock markets. The well-known external shocks, such as the COVID-19 pandemic-induced market instability, the mass hysteria, and the corporate debt bubble, only served as sparks during the 2020 stock market crash, but they were not the root causes of the 2020 U.S. stock market crash.

From a historic point of view, between 1954 and 2009, the U.S. Stock market have gone through many cycles of boom and bust with an average length of boom (expansion) period of 58 months, 2 months shy of 5 years, and the bust period averaged at 11 months. The longest boom before 2009 occurred between March 1991 and March 2001, lasting for a total of 10 years. This record 10-year boom is only eclipsed by the most recent 11-year boom from 2009 to 2020 until the latter was interrupted by the COVID crash. Statistically, as confirmed by our LPPLS analysis, the market was due to have a crash internally, and the COVID-19 pandemic only served as an external trigger, hence the 2020 U.S. stock market crash might be more suitably referred to as the 'COVID' crash, rather than the Great COVID crash.

Furthermore, we also performed the complementary post-mortem analysis to further characterize the estimated LPPLS models for detecting the 2020 U.S. stock market crash. The analyses indicate that the probability density distributions of the predicted critical time for these four indexes are positively skewed, and the 2020 U.S. stock market crash originates from a bubble which began to form as early as September 2018. In addition, the SP400 index and R2000 index have the distinctly different probability density distribution shapes compared to the W5000 and SP500 indexes, indicating that the bubbles in stocks with different levels of total market capitalizations may have significantly different starting time profiles.

This study not only provides a novel dissection of the 2020 stock market crash in the United States, but also creates a paradigm for future studies in real-time crash detection and underlying mechanism classification. It serves to warn us of the imminent risks in the stock market as well as other financial markets and economic indexes.

## Acknowledgment

The work is supported by the Faculty Research Initiative Grant as well as the New Faculty Start-Up Funds at the University of Wisconsin-Stout. The authors would like to thank the Blugold Supercomputing Cluster (BGSC) at the University of Wisconsin-Eau Claire.